\documentclass[11pt]{article}

\usepackage[final]{acl}

\usepackage{times}
\usepackage{latexsym}
\usepackage{multirow}

\usepackage[T1]{fontenc}

\usepackage[utf8]{inputenc}

\usepackage{microtype}

\usepackage{inconsolata}

\usepackage{graphicx}
\usepackage{tabularx}

\usepackage{amsmath}

\usepackage{booktabs}
\usepackage[table]{xcolor}

\usepackage{booktabs}
\usepackage{graphicx}
\usepackage{float}

\usepackage{booktabs}
\usepackage{longtable}
\usepackage{array}
\usepackage{ragged2e}
\usepackage{caption}

\usepackage{booktabs}
\usepackage{multirow}
\usepackage{tabularx}
\usepackage{array}
\usepackage[table]{xcolor}
\usepackage{pifont}

\newcommand{\cmark}{\ding{51}}

\usepackage{hyperref}
\usepackage{url} 

\usepackage{booktabs}
\usepackage{multirow}
\usepackage[table]{xcolor}

\usepackage[most]{tcolorbox}
\usepackage{xcolor}
\definecolor{prompttitlebg}{HTML}{75999B} 
\definecolor{promptbodybg}{HTML}{F1F6F7}  

\usepackage[table]{xcolor}

\definecolor{bestcolor}{HTML}{FAF0C8}      
\definecolor{secondcolor}{HTML}{DCEBFA}    
\definecolor{grouptext}{HTML}{666666}

\newcommand{\bestcell}[1]{\cellcolor{bestcolor}\textbf{#1}}
\newcommand{\secondcell}[1]{\cellcolor{secondcolor}\underline{#1}}

\newtcolorbox{promptbox}[1]{
    enhanced jigsaw, 
    colframe=prompttitlebg,
    colback=promptbodybg,
    coltitle=white,
    fonttitle=\bfseries,
    title=#1,
    arc=2mm,
    boxrule=0.5pt,
    left=3mm, right=3mm, top=3mm, bottom=3mm, 
    toptitle=1.5mm, 
    bottomtitle=1.5mm, 
    lefttitle=3mm, 
    fontupper=\small,
    boxsep=0pt,
    before skip=1em,
    after skip=1em,
    toprule at break=0pt,
    bottomrule at break=0pt,
    pad at break=0mm
}

\title{SemFlowRAG: Directed Semantic Flow from Abstraction to Evidence \\ for Complex Reasoning}

\author{
  Houyuan Qin$^{1}$\thanks{Equal contribution.},
  Rong Wu$^{1,2}$\footnotemark[1],
  Qinyuan Qin$^{1}$,
  Botian Shi$^{1}$
  \\
  {\bf
  Jingjing Qu$^{1}$,
  Yang Sun$^{1,3}$,
  Pinlong Cai$^{1}$\thanks{Corresponding author: caipinlong@pjlab.org.cn}}
  \\
  \\
  $^{1}$Shanghai Artificial Intelligence Laboratory \\
  $^{2}$Zhejiang University \\
  $^{3}$Fudan University
}

\begin{document}
\maketitle
\begin{abstract}
Retrieval-Augmented Generation (RAG) enhanced by Knowledge Graphs has shown promise in complex multi-hop reasoning tasks. However, existing graph-based retrieval methods typically rely on flat, undirected topologies. During the retrieval process, the probability flow often gets trapped in high-degree abstract concept nodes which we define as ``probability black holes'', leading to semantic drift and noise accumulation. To address this, we propose SemFlowRAG, a framework that reconstructs the flat retrieval space into a corpus-adaptive semantic gradient graph. This data-driven self-organization enables a hierarchical structure to emerge naturally from the data distribution, capturing the intrinsic semantic granularity of the corpus to suppress structural noise. By quantifying the semantic abstractness of entities through the embedding variance of their associated passages, we transform static undirected edges into directed semantic constraints. Furthermore, we design an abstractness-guided directed PageRank algorithm that forces the retrieval trajectory to follow a ``high-to-low semantic abstractness'' gradient. This mechanism ensures layer-by-layer evidence convergence, smoothly guiding the retrieval process from abstract concepts to specific document evidence. Extensive experiments on complex QA datasets demonstrate that SemFlowRAG effectively mitigates the ``probability black holes'' issue, outperforming existing baselines in both retrieval and downstream reasoning performance. Code is available at \url{https://github.com/KnowledgeXLab/SemFlowRAG}.

\end{abstract}

\section{Introduction}

Large Language Models (LLMs) have demonstrated remarkable performance across diverse natural language processing tasks~\cite{achiam2023gpt, yang2025qwen3}. However, they remain constrained by several inherent limitations, such as hallucination and the inability to access up-to-date or domain-specific knowledge~\cite{huang2025survey, ji2023towards}. Retrieval-Augmented Generation (RAG) effectively mitigates these issues by integrating external corpora and performing dynamic retrieval during the inference phase~\cite{fan2024survey, arslan2024survey}. However, while vector-based RAG excels at simple factual queries, it struggles with complex reasoning tasks like multi-hop question answering, as it retrieves isolated semantic snippets that lack the coherent evidence chains and cross-document logical dependencies~\cite{fan2024survey, wu2025kg}. Consequently, constructing semantically structured knowledge networks from large-scale document corpora to establish systematic knowledge topologies has emerged as a critical research direction for overcoming the current performance bottlenecks in RAG systems.

To bridge this gap, Graph-based RAG converts unstructured corpora into Knowledge Graph (KG) with explicit logical relations, enabling retrieval for reasoning~\cite{edge2024local, zhang2026leanrag}. However, current approaches mainly generate flat and static topological spaces, lacking fine-grained semantic constraints in both knowledge organization and path retrieval~\cite{han2025rag, han2024retrieval}. This oversight manifests in two critical limitations:

First, at the knowledge construction level, self-organizing mechanisms are prone to semantic construction bias. Predominant approaches rely on community detection (e.g., Microsoft GraphRAG, LightRAG)~\cite{edge2024local, guo2024lightrag} or leverage LLM-driven summarization for hierarchical modeling (e.g., RAPTOR, LeanRAG)~\cite{raptor,zhang2026leanrag}. However, these strategies over-rely on generic model priors for information aggregation, failing to capture the intrinsic logical structure of domain-specific corpora.

Second, at the query retrieval level, mainstream strategies employ semantic hierarchy-agnostic random-walk traversal (e.g., HippoRAG)~\cite{hipporag}. Without semantic hierarchical guidance, the walker drifts toward high-centrality abstract nodes~\cite{he2024acquiring,veremyev2019graph}, which absorb a disproportionate share of retrieval probability at the expense of fine-grained evidence. We term this the \textit{``Probability Black Hole.''} 
Consequently, ensuring accurate, traceable multi-hop reasoning requires semantic hierarchy-aware structure modeling and retrieval alignment.


\begin{figure*}[h]
    \centering
    \includegraphics[width=0.8
    \linewidth]{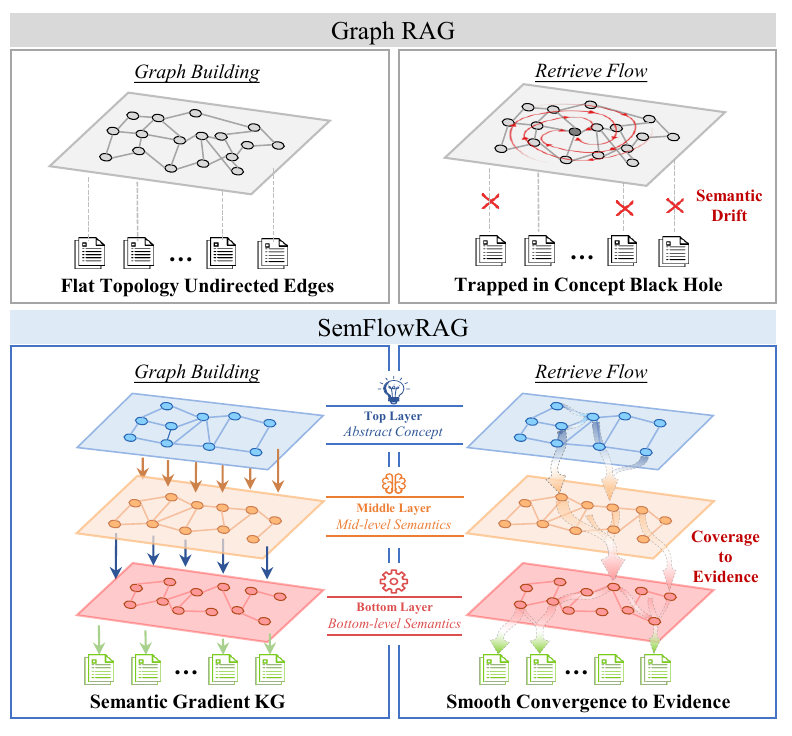}
    \caption{\textbf{Conceptual comparison.} (Top) Standard Graph RAG uses flat, undirected graphs; random walks suffer from ``probability black holes'' at high-degree abstract nodes. (Bottom) SemFlowRAG enforces top-down directed flow along semantic gradients, routing probability from abstract concepts to specific evidence.}
    \label{fig:teaser}
\end{figure*}

To address these limitations, we propose SemFlowRAG, which constructs a hierarchical semantic gradient graph in a data-driven manner (Figure~\ref{fig:teaser}). The abstractness of each entity is quantified by the embedding variance of its associated passages, allowing the knowledge hierarchy to emerge from the corpus itself without external annotation. We then orient edges top-down along this gradient, establishing directed semantic information flow for complex reasoning.

Tailored to this structure, we introduce a directed Personalized PageRank (PPR) retrieval paradigm with a smooth penalty mechanism that routes probability mass downward (from abstract concepts to concrete evidence) preventing probability trapping at intermediate nodes. A preliminary analysis confirms this: our directed routing reduces top-10 entity abstractness by 73.6\% (0.321 $\rightarrow$ 0.085) and boosts query similarity in ranks 60-100 by 11.3\%. Unlike macro-level Graph RAG methods~\cite{edge2024local,luo2024reasoning} that rely on discrete structural clustering and LLM-generated community summaries, SemFlowRAG injects a continuous semantic gradient into the KG, guiding retrieval to naturally converge on specific evidence without explicit graph clustering. Extensive experiments on multiple complex QA benchmarks demonstrate significant gains over existing Graph RAG baselines.

In summary, main contributions are as follows:
\begin{itemize}
    \vspace{-0.25em}
    \item We identify the ``probability black hole'' issue in existing flat Graph RAG paradigms and propose a dynamic graph reconstruction method that transforms static topologies into semantic gradient graphs based on entity semantic abstractness.
    \vspace{-0.25em}
    \item We design a semantic abstractness-guided directed PPR algorithm equipped with a smooth penalty mechanism to navigate the semantic gradient graph. By constraining the retrieval flow along the ``high-to-low abstraction'' direction, the algorithm ensures layer-by-layer evidence convergence.
    \vspace{-0.25em}
    \item We conduct extensive experiments across multiple complex QA benchmarks. Empirical results demonstrate that SemFlowRAG outperforms existing Graph RAG baselines, validating the effectiveness of integrating semantic gradients into structural retrieval.
\end{itemize}

\section{Related Work}
\label{sec:related_work}

\subsection{Structured Knowledge Representation}
\label{subsec:related_structure}

Early RAG systems typically organize documents into flat text chunks and rely on embedding-based similarity for retrieval~\cite{lewis2020retrieval}. To better capture inter-document relationships, subsequent works have explored graph-structured knowledge representations. For instance, GraphRAG~\cite{edge2024local} organizes text into community-based knowledge graphs, while HippoRAG~\cite{hipporag} constructs personalized page graphs to simulate associative memory during retrieval. These methods represent a meaningful step beyond flat chunk retrieval by explicitly encoding relational structure.

To further enable multi-granularity access to knowledge, several recent approaches have introduced hierarchical organization. RAPTOR~\cite{raptor} builds a tree structure via recursive clustering and summarization of text chunks. GraphRAG~\cite{edge2024local} generates community summaries at multiple levels, effectively creating a multi-resolution view of the corpus. HiRAG~\cite{jiao2025hirag}, LightRAG~\cite{guo2024lightrag} and LeanRAG~\cite{zhang2026leanrag} similarly adopt hierarchical architectures to support retrieval at varying levels of abstraction. However, these methods primarily treat the hierarchy as a multi-granularity index: they organize knowledge into layers of increasing abstraction but do not explicitly model the \textit{semantic gradient} between abstract concepts and concrete evidence. In other words, they lack a principled mechanism to quantify how much semantic abstractness a given node contains or to distinguish abstract, high-frequency concepts from specific, low-frequency details. Our work addresses this gap by introducing abstractness-based hierarchy modeling, which provides an explicit structural prior to differentiate abstract and concrete knowledge during retrieval.

\subsection{Graph-based RAG}
\label{subsec:graph_rag}

Beyond knowledge representation, how to effectively retrieve the constructed graph to get relevant evidence is a central challenge in GraphRAG. Existing graph retrieval approaches mainly include two paradigms: GNN-based message passing and random-walk-based diffusion. GNN-RAG~\cite{mavromatis2025gnn} employs graph neural networks to propagate information across neighboring nodes, producing context-aware representations for downstream generation. While effective for localized reasoning, the receptive field of GNNs is constrained by the number of message-passing layers, and training such models introduces extra computational overhead. More recently, HippoRAG~\cite{hipporag} adopted PPR~\cite{haveliwala2002topic} to simulate human-like associative retrieval, using undirected random walks over personalized page graphs to identify evidence nodes in an unsupervised manner.

Despite their effectiveness, both paradigms share a common limitation: they treat the retrieval graph as undirected and overlook difference in semantic hierarchy. During diffusion, the walker's transition is governed primarily by structural connectivity and local semantic similarity, with no mechanism to distinguish abstract, high-centrality concept nodes from specific evidence nodes. In densely connected knowledge graphs, this causes probability mass to accumulate at high-degree concept nodes leading to semantic drift and diluting the retrieval signal for fine-grained evidence passages. To address this, we propose a top-down directed routing mechanism. By quantifying node-level semantic abstractness and incorporating an abstractness penalty into the transition weights, we explicitly regularize the random walk to flow from abstract concepts toward specific evidence passages, effectively mitigating probability trapping at intermediate, non-evidential nodes.

\section{Methodology}
\vspace{-0.25em}
\subsection{Overview}

\begin{figure*}[h]
    \centering
    \includegraphics[width=0.95
    \linewidth]{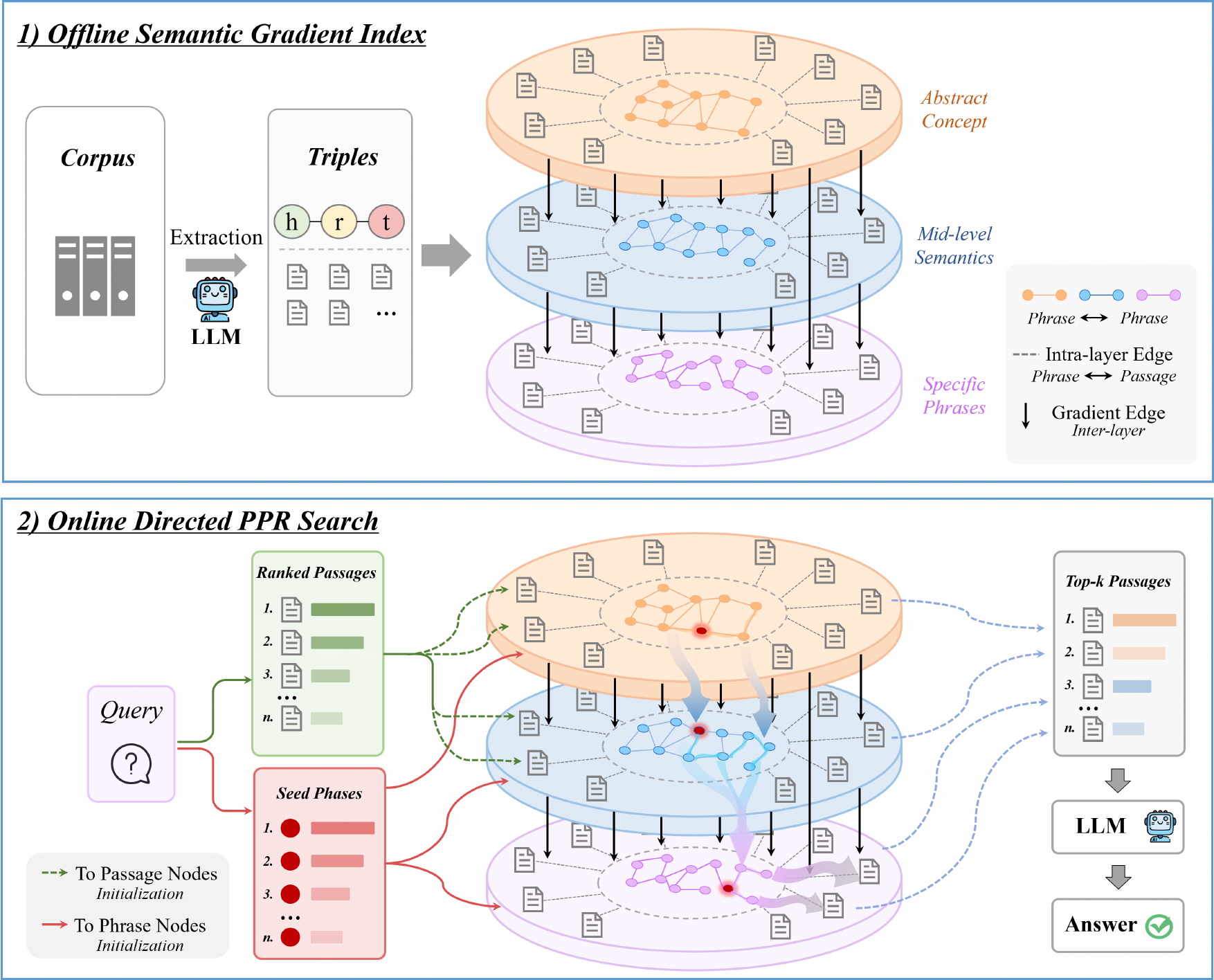}
    \caption{\textbf{Overview of SemFlowRAG.} \textbf{1) Offline:} We construct a hierarchical graph with top-down directed edges, stratifying nodes by semantic abstractness. \textbf{2) Online:} A directed PPR search routes probability mass from query seed nodes downward along semantic gradients to retrieve fine-grained evidence passages.}
    \vspace{-0.75em}
\label{fig:framework}
\end{figure*}

The proposed SemFlowRAG framework operates in a two-stage paradigm: offline graph construction and online directed retrieval, as illustrated in Figure~\ref{fig:framework}. In the offline stage, we upgrade a standard flat KG into a \textit{Semantic Gradient KG} by explicitly quantifying the semantic abstractness of entities based on their associated passages. This step transforms undirected, static KG into a knowledge representation with an inherent top-down semantic hierarchy. Built upon this representation, the online stage introduces a semantic information-guided directed retrieval mechanism. By employing a directionally constrained PPR, the retrieval trajectory is explicitly guided to follow the semantic gradient, smoothly penetrating from generalized concepts and ultimately converging on the passage nodes that contain concrete and relevant evidence.

\subsection{Offline Stage: Semantic Gradient KG Construction}
\label{sec:offline_stage}

The primary objective of offline stage is to reconstruct a static, flat KG into a knowledge representation network with an intrinsic semantic hierarchy. This is achieved through three sequential steps: basic graph extraction, semantic abstractness quantification and semantic gradient edge construction.

\paragraph{Entity and Fact Extraction.}
Given a corpus of documents, we first segment them into retrievable passages and cache their dense embeddings. For each passage, we apply Open Information Extraction (OpenIE) to extract named entities and factual relational triplets (i.e., subject-relation-object). The extracted subjects and objects are normalized and mapped to entity nodes in the graph. We establish unweighted \textit{entity-entity} edges based on the co-occurrence of entities within the same triplet. Simultaneously, to preserve the source of the knowledge, we record the occurrences of each entity and construct bidirectional \textit{passage-entity} edges, explicitly linking entities to the specific passages they originate from.

\paragraph{Semantic Abstractness Quantification.}
In existing flat KGs~\cite{liang2025kag, jiao2025hirag, hipporag2}, all entities are treated as topologically equivalent, ignoring the semantic hierarchical distinction between broad and specific concepts. To explicitly model this hierarchy, we propose a metric to quantify the ``semantic abstractness'' of each entity. 
Let $\mathcal{P}(v)$ denote the set of passages containing entity $v$, and $\mathbf{h}_p$ be the dense embedding of passage $p$. We define the semantic abstractness $d_v$ of entity $v$ as the trace of the covariance matrix of its associated passage embeddings:
\begin{equation}
    d_v = \mathrm{tr}\left(\mathrm{Var}\left(\{\mathbf{h}_p \mid p\in\mathcal{P}(v)\}\right)\right).
\end{equation}
Mathematically, the covariance matrix $\mathrm{Var}(\cdot)$ captures the dispersion of the entity's contextual embeddings across the semantic space. Taking the trace $\mathrm{tr}(\cdot)$ sums the variances along all principal dimensions, effectively computing the total spatial variance. The fundamental insight behind this formulation is that abstract, high-level concepts (e.g., ``country'', ``science'') naturally appear in highly diverse and semantically disconnected contexts, resulting in a widely scattered distribution and a large total variance. Conversely, specific, fine-grained entities (e.g., ``the 2024 Paris Olympics'') are typically constrained to a narrow, well-defined semantic region, yielding a highly concentrated distribution and a small variance. Thus, $d_v$ serves as an unsupervised proxy for corpus-level semantic breadth, rather than universal ontological abstractness.

To mitigate the influence of extreme outliers, we apply a truncated min-max normalization to the abstractness scores of all entities, clipping the values at the $1^{\text{st}}$ and $99^{\text{th}}$ percentiles. This scales $d_v$ into a standardized range of $[0, 1]$.

\paragraph{Semantic Gradient Edge Construction.}
Based on the quantified semantic abstractness, we establish a semantic gradient across the graph. Instead of treating the entity-entity connections as purely undirected co-occurrence links, we compute the semantic span for every connected pair of entities $u$ and $v$ as $\Delta_{u,v} = d_v - d_u$. 

By explicitly recording semantic gradient priors ($\Delta_{u,v}$) offline, we transform the flat topological space into a directed, hierarchical knowledge representation. This offline gradient builds a ``high-to-low abstraction'' scaffold within the graph, which serves as the fundamental constraint for guiding the online probability flow and preventing the retrieval trajectory from converging into abstract nodes.

\subsection{Online Stage: Semantic Information-Guided Directed Retrieval}
\label{sec:online_stage}

With the Semantic Gradient KG constructed offline, the online stage aims to navigate this hierarchical topology based on a given query. We formulate this process as a dynamically weighted, directed PPR algorithm, which explicitly forces the retrieval trajectory to smoothly converge upon fine-grained and relevant evidence.

\paragraph{Seed Retrieval and Initialization.}
Given a query $q$, the system first identifies the initial entry points (seeds) for the graph random walk. We compute the similarity between $q$ and all extracted facts to retrieve the top-$k$ candidate facts. An optional LLM-based filter can be applied to retain only high-quality, query-relevant facts. If no facts pass the filter, the system falls back to standard Dense Passage Retrieval (DPR) without graph propagation.

For the retained facts, their subjects and objects are mapped to the corresponding entity nodes to serve as seeds. To mitigate the topological bias of high-frequency entities, the initial weight of each seed entity is determined by its query-fact similarity normalized by its occurrence frequency in the corpus. Furthermore, we incorporate the DPR scores of passage nodes as a weak prior, controlled by a weight coefficient. Consequently, the final PPR reset distribution (the restart vector) is a hybrid formulation comprising both the normalized entity seed weights and the passage prior weights. More details are provided in the Appendix~\ref{appendix:implementation_datails}.

\paragraph{Semantic Information-Guided Layer-by-Layer Convergence.}
The core of our directed retrieval lies in the dynamic rewriting of transition probabilities. During the random walk, the algorithm dynamically balances the online \textit{query relevance} with the offline \textit{semantic gradient}. For entity $v$, we define its query relevance as the maximum similarity between $q$ and its associated passages:
\begin{equation}
    Rel(v|q) = \max_{p \in \mathcal{P}(v)} \text{sim}(q, p),
\end{equation}
which is min-max normalized to $[0,1]$. A higher $Rel(v|q)$ indicates that traversing through $v$ is more likely to yield query-relevant contexts.

To determine the transition score $s(u,v)$ for a directed edge $u \rightarrow v$, we inject a smooth penalty mechanism:
\begin{equation}
    \label{edge_weight}
    s(u,v) = Rel(v \mid q) - \lambda |\Delta_{u,v}|,
\end{equation}
where $\Delta_{u,v} = d_v - d_u$ is the pre-computed semantic gradient from the offline stage, and $\lambda$ is a penalty coefficient, we set $\lambda=1$ as default. This formulation explicitly rewards paths leading to query-relevant nodes while penalizing excessively large jumps in semantic abstractness. By penalizing large semantic gaps, we enforce a smooth gradient, ensuring that the retrieval trajectory follows a step-by-step reasoning chain and mitigating the risk of topic drift during the random walk.

To enforce the top-down semantic information flow, we apply an asymmetric directional routing strategy to the graph edges. Specifically, out of the probability mass allocated for edge transitions, the vast majority (e.g., $90\%$) is distributed to \textit{down-edges} (transitions from higher to lower semantic abstractness), while a marginal fraction (e.g., $10\%$) is reserved for \textit{up-edges}. This design ensures that the probability mass smoothly penetrates downwards layer by layer, effectively avoiding the ``probability black holes'' associated with high-frequency abstract nodes. Meanwhile, we preserve a small amount of upward connectivity to simulate the associative recall process in human memory, where reasoning does not proceed strictly downward but occasionally returns to broader concepts. Note that passage-entity edges are kept bidirectional and unweighted to allow flexible information interaction between entity nodes and passage nodes.

\vspace{-0.5em}
\paragraph{Passage Ranking and Output.}
Driven by the directionally constrained PPR, the probability flow trickles down the semantic gradient and naturally accumulates on the specific passage nodes. Once the PPR reaches its steady-state distribution, we directly extract the PageRank scores of all passage nodes. The top-$k$ scoring passages are retrieved as the final supporting context, which is subsequently fed into the downstream LLM to generate answer.

\begin{table*}[!t]
\centering
\small
\begin{tabular*}{\textwidth}{@{\extracolsep{\fill}}lccccccc@{}}
\toprule
 & \textbf{NQ} & \textbf{PopQA} & \textbf{MuSiQue} & \textbf{2Wiki} & \textbf{HotpotQA} & \textbf{LV-Eval} & \textbf{NarrativeQA} \\
\midrule
\textbf{Num of queries}  & 1,000 & 1,000 & 1,000 & 1,000 & 1,000 & 124 & 293 \\
\textbf{Num of passages} & 9,633 & 8,676 & 11,656 & 6,119 & 9,811 & 22,849 & 4,111 \\
\bottomrule
\end{tabular*}
\vspace{-1em}
\caption{Dataset statistics used in our evaluation.}

\label{tab:dataset-statistics}
\end{table*}

\section{Experiments}
\label{sec:experiments}
In this section, we empirically evaluate SemFlowRAG. Our experiments are designed to answer the following core research questions (RQs):

\begin{itemize}
    \vspace{-0.5em}
    \item \textbf{RQ1 (Main Results):} Can SemFlowRAG achieve state-of-the-art performance on complex multi-hop reasoning tasks compared to existing RAG baselines?
    \vspace{-0.5em}
    \item \textbf{RQ2 (Ablation on Topology):} Does enforcing a directed walk along the semantic gradient improve performance compared to standard undirected traversal?
    \vspace{-0.5em}
    \item \textbf{RQ3 (Ablation on Weighting):} Is the abstractness penalty mechanism in the transition probability more effective than relying solely on query similarity?
\end{itemize}

\subsection{Experimental Setup}
\label{subsec:setup}
\textbf{Datasets.} We evaluate SemFlowRAG on a diverse suite of question answering benchmarks covering factual retrieval, open-domain QA, multi-hop reasoning and long-context evidence aggregation, including NaturalQuestions (NQ)~\cite{kwiatkowski2019natural}, PopQA~\cite{mallen2023not}, MuSiQue~\cite{trivedi2022musique}, 2WikiMultiHopQA~\cite{ho2020constructing}, HotpotQA~\cite{yang2018hotpotqa}, LV-Eval~\cite{yuan2024lv} and NarrativeQA~\cite{kovcisky2018narrativeqa}. 
These datasets collectively test whether a system can move beyond shallow semantic matching and identify evidence chains that support complex reasoning. We evaluate each dataset on a selected subset of queries, with the corresponding dataset statistics summarized in Table~\ref{tab:dataset-statistics}.

\textbf{Baselines.} We compare SemFlowRAG with three groups of baselines. First, we include unstructured knowledge representation augmented baselines, covering a no-retrieval setting for QA, the sparse retriever BM25~\cite{robertson1994some} and so on. Second, we compare embedding-based knowledge representation augmented baselines, including GTE-Qwen2-7B-Instruct~\cite{li2023towards}, GritLM-7B~\cite{muennighoff2025generative} and so on. Third, we include structured knowledge representation augmented methods, including RAPTOR~\cite{raptor}, HippoRAG 2~\cite{hipporag2} and so on. Baseline details are provided in the Appendix~\ref{appendix:baseline_details}. For fair comparison, reproduced structure-augmented baselines are evaluated under the same LLM, embedding model, retrieval budget and generation setting whenever applicable. Additional implementation details are provided in the Appendix~\ref{appendix:implementation_datails}.

\textbf{Metrics.} For retrieval, we report Passage Recall@5, defined as the fraction of gold evidence passages successfully retrieved within the top-5 results. For answer generation, we compute F1 based on exact match (EM) between the generated answers and the ground-truth answers. All results are reported under identical top-5 retrieval and generation settings to ensure fair comparison.

\begin{table*}[!t]
\centering
\small
\setlength{\tabcolsep}{4.0pt}
\renewcommand{\arraystretch}{1.05}

\begin{tabularx}{\textwidth}{@{}l *{8}{>{\centering\arraybackslash}X}@{}}
\toprule
\multirow{2}{*}{Method} 
& \multicolumn{2}{c}{Simple QA} 
& \multicolumn{4}{c}{Multi-Hop QA} 
& \multicolumn{1}{c}{Discourse} 
& \multirow{2}{*}{Avg.} \\
\cmidrule(lr){2-3} \cmidrule(lr){4-7} \cmidrule(lr){8-8}
& NQ & PopQA & MuSiQue & 2Wiki & HotpotQA & LV-Eval & NarrativeQA & \\
\midrule

\multicolumn{9}{@{}l}{\textcolor{grouptext}{\textit{Unstructured Knowledge Representation}}} \\
Llama-3.3-70B-Instruct w/o Retrieval
& 54.9 & 32.5 & 26.1 & 42.8 & 47.3 & 6.0 & 12.9 & 38.4 \\
BM25 
& 59.0 & 49.9 & 28.8 & 51.2 & 63.4 & 5.9 & 18.3 & 47.7 \\
Contriever 
& 58.9 & 53.1 & 31.3 & 41.9 & 62.3 & 8.1 & 19.7 & 46.9 \\
GTR 
& 59.9 & \secondcell{56.2} & 34.6 & 52.8 & 62.8 & 7.1 & 19.9 & 50.4 \\

\midrule
\multicolumn{9}{@{}l}{\textcolor{grouptext}{\textit{Embedding-based Knowledge Representation}}} \\
GTE-Qwen2-7B-Instruct 
& 62.0 & \bestcell{56.3} & 40.9 & 60.0 & 71.0 & 7.1 & 21.3 & 54.9 \\
GritLM-7B 
& 61.3 & 55.8 & 44.8 & 60.6 & 73.3 & 9.8 & 23.9 & 56.1 \\
NV-Embed-v2 (7B) 
& 61.9 & 55.7 & 45.7 & 61.5 & 75.3 & 9.8 & 25.7 & 57.0 \\

\midrule
\multicolumn{9}{@{}l}{\textcolor{grouptext}{\textit{Structured Knowledge Representation}}} \\
RAPTOR 
& 50.7 & \secondcell{56.2} & 28.9 & 52.1 & 69.5 & 5.0 & 21.4 & 48.8 \\
GraphRAG 
& 46.9 & 48.1 & 38.5 & 58.6 & 68.6 & \secondcell{11.2} & 23.0 & 49.6 \\
LightRAG 
& 16.6 & 2.4 & 1.6 & 11.6 & 2.4 & 1.0 & 3.7 & 6.6 \\
HippoRAG 
& 55.3 & 55.9 & 35.1 & 71.8 & 63.5 & 8.4 & 16.3 & 53.1 \\
HippoRAG 2 
& \bestcell{63.3} & \secondcell{56.2} & \secondcell{48.6} & 71.0 & \secondcell{75.5} & \bestcell{12.9} & 25.9 & \secondcell{59.8} \\
HippoRAG 2 (reproduced) 
& 62.1 & 55.7 & \secondcell{48.6} & \secondcell{72.5} & \secondcell{75.5} & 9.6 & \secondcell{26.2} & 59.6 \\
\midrule
\textbf{SemFlowRAG (Ours)} 
& \secondcell{62.3} & \secondcell{56.2} & \bestcell{51.4} & \bestcell{76.9} & \bestcell{75.8} & 10.6 & \bestcell{26.6} & \bestcell{61.2} \\

\bottomrule
\end{tabularx}
\vspace{-3mm}
\caption{
Reasoning performance (F1 score) on seven benchmarks.
\protect\colorbox{bestcolor}{\textbf{Best}} and
\protect\colorbox{secondcolor}{\underline{second-best}} are highlighted.
}
\vspace{-3mm}
\label{tab:main_results}
\end{table*}

\subsection{Main Results (RQ1)}
\label{subsec:main_results}

Table~\ref{tab:main_results} and Table~\ref{tab:retrieval_results} present the downstream QA and evidence retrieval results. SemFlowRAG achieves the highest overall performance, yielding an average F1 of 61.2 and Recall@5 of 79.3. 

The gains are most pronounced on multi-hop reasoning tasks (MuSiQue, 2Wiki and HotpotQA). SemFlowRAG outperforms the strongest baseline, HippoRAG 2, by 4.4 and 2.8 F1 points on 2Wiki and MuSiQue, respectively, while reaching a 94.8 Recall@5 on 2Wiki. This validates that constraining retrieval along a semantic gradient provides reliable evidence paths for complex queries, effectively mitigating noise accumulation and semantic drift. On Simple QA (NQ, PopQA) and Discourse (NarrativeQA) tasks, SemFlowRAG remains highly competitive (e.g., achieving the highest F1 of 26.6 on NarrativeQA). This demonstrates that our design aids long-context synthesis without degrading single-hop retrieval efficiency.

Notably, while standard undirected or tree-based methods (e.g., GraphRAG, RAPTOR) sometimes underperform the dense retriever (NV-Embed-v2) due to graph-induced structural noise, SemFlowRAG consistently improves upon NV-Embed-v2. This confirms that our directed edges effectively filter noise and accurately guide the reasoning trajectory.

\begin{table*}[!t]
\centering
\small
\setlength{\tabcolsep}{4.0pt}
\renewcommand{\arraystretch}{1.05}

\begin{tabularx}{\textwidth}{@{}l *{6}{>{\centering\arraybackslash}X}@{}}
\toprule
\multirow{2}{*}{Method} 
& \multicolumn{2}{c}{Simple QA} 
& \multicolumn{3}{c}{Multi-Hop QA} 
& \multirow{2}{*}{Avg.} \\
\cmidrule(lr){2-3} \cmidrule(lr){4-6}
& NQ & PopQA & MuSiQue & 2Wiki & HotpotQA & \\
\midrule

\multicolumn{7}{@{}l}{\textcolor{grouptext}{\textit{Unstructured Knowledge Representation Augmented}}} \\
BM25 
& 56.1 & 35.7 & 43.5 & 65.3 & 74.8 & 55.1 \\
Contriever 
& 54.6 & 43.2 & 46.6 & 57.5 & 75.3 & 55.4 \\
GTR 
& 63.4 & 49.4 & 49.1 & 67.9 & 73.9 & 60.7 \\

\midrule
\multicolumn{7}{@{}l}{\textcolor{grouptext}{\textit{Embedding-based Knowledge Representation Augmented}}} \\
GTE-Qwen2-7B-Instruct 
& 74.3 & 50.6 & 63.6 & 74.8 & 89.1 & 70.5 \\
GritLM-7B 
& 76.6 & 50.1 & 65.9 & 76.0 & 92.4 & 72.2 \\
NV-Embed-v2 (7B) 
& 75.4 & 51.0 & 69.7 & 76.5 & 94.5 & 73.4 \\

\midrule
\multicolumn{7}{@{}l}{\textcolor{grouptext}{\textit{Structured Knowledge Representation Augmented}}} \\
RAPTOR 
& 68.3 & 48.7 & 57.8 & 66.2 & 86.9 & 65.6 \\
HippoRAG (reproduced) 
& 44.4 & \bestcell{53.8} & 53.2 & 90.4 & 77.3 & 63.8 \\
HippoRAG 2 
& \bestcell{78.0} & 51.7 & \secondcell{74.7} & 90.4 & \bestcell{96.3} & \secondcell{78.2} \\
HippoRAG 2 (reproduced) 
& 77.0 & 51.7 & 73.8 & \secondcell{90.7} & \bestcell{96.3} & 77.9 \\
\midrule
\textbf{SemFlowRAG (Ours)} 
& \secondcell{77.5} & \secondcell{52.8} & \bestcell{75.2} & \bestcell{94.8} & \secondcell{96.1} & \bestcell{79.3} \\

\bottomrule
\end{tabularx}

\caption{
Retrieval performance (passage recall@5) on five benchmarks.
\protect\colorbox{bestcolor}{\textbf{Best}} and
\protect\colorbox{secondcolor}{\underline{second-best}} are highlighted.
}
\label{tab:retrieval_results}
\end{table*}

\subsection{Ablation Study}
\label{subsec:ablation}
To understand the contribution of each component in SemFlowRAG, we conduct careful ablation studies focusing on the graph topology and the transition weights. Table~\ref{tab:ablation} presents the ablation results of SemFlowRAG on multi-hop QA benchmarks. We evaluate the contribution of direction control, query similarity and abstractness difference in semantic-aware graph retrieval.

\begin{table*}[!t]
\centering
\small
\setlength{\tabcolsep}{5.0pt}
\renewcommand{\arraystretch}{1.08}

\begin{tabularx}{\textwidth}{@{}lccc *{4}{>{\centering\arraybackslash}X}@{}}
\toprule
\multirow{2}{*}{Method}
& \multicolumn{3}{c}{Components}
& \multicolumn{4}{c}{Multi-Hop QA} \\
\cmidrule(lr){2-4} \cmidrule(lr){5-8}
& Direction Control & Query Similarity & Delta Abstractness
& MuSiQue & 2Wiki & HotpotQA & Avg. \\
\midrule

SemFlowRAG 
& \cmark & \cmark & \cmark
& \textbf{51.4} & \textbf{76.9} & \textbf{75.8} & \textbf{68.0} \\

\textit{w/o Direction Control}
&  & \cmark & \cmark
& 50.8 & 76.0 & 75.5 & 67.4 \\

\textit{w/o Query Similarity}
& \cmark &  & \cmark
& 50.9 & 76.5 & 75.3 & 67.6 \\

\textit{w/o Delta Abstractness}
& \cmark & \cmark & 
& 51.2 & 76.4 & 75.3 & 67.6 \\

\bottomrule
\end{tabularx}

\caption{
Ablation study of SemFlowRAG on multi-hop QA benchmarks.
}
\vspace{-1.5em}
\label{tab:ablation}
\end{table*}

\subsubsection{Effect of Directed Topology (RQ2)}

To answer RQ2, we evaluate the ``w/o Direction Control'' variant by disabling the asymmetric directional retrieval. In this setting, instead of forcing the probability mass to flow primarily downwards (from high to low), we allow it to distribute symmetrically across both down-edges and up-edges.

As shown in Table~\ref{tab:ablation}, removing this directional constraint reduces the average F1 score from 68.0 to 67.4. Without the constraint, the transition probabilities rely primarily on the edge weight $s(u,v)$. Because the abstractness penalty $|\Delta_{u,v}|$ in the edge weight is symmetric, the random walk can easily move upward toward broader concepts that also exhibit high query relevance. Consequently, the retrieval trajectory tends to wander among abstract entities instead of actively penetrating down to the specific passages where the actual evidence resides. The performance drop is most notable on 2Wiki, confirming that for scenarios requiring longer reasoning chains, an explicit downward constraint is necessary to prevent semantic drift and ensure the walk converges on concrete evidence.

\subsubsection{Effect of Semantic Weighting (RQ3)}

To answer RQ3, we evaluate the two components that formulate the transition weights in Equation~\ref{edge_weight}: query similarity and the abstractness penalty. In the ``w/o Query Similarity'' variant, we remove the relevance score $Rel(v \mid q)$, making the transition probabilities depend solely on the abstractness difference between nodes. Conversely, in the ``w/o Delta (abstractness)'' variant, we set the penalty coefficient $\lambda = 0$, relying entirely on the semantic similarity between the query and the target node to guide the random walk. 

As shown in Table~\ref{tab:ablation}, removing either component reduces the average F1 score. Without query similarity, the random walk blindly follows the structural gradient to lower-level nodes, risking the retrieval of evidence that is structurally reachable but irrelevant to the initial question. On the other hand, removing the abstractness penalty causes the walk to rely only on local textual similarity. This can lead to abrupt semantic jumps to nodes that are textually similar to the query but disrupt the continuous, smooth retrieval chain. The equal performance drop indicates that both signals are complementary and indispensable: query similarity ensures task-specific relevance, while the abstractness penalty acts as a regularization term to maintain smooth transitions along the semantic hierarchy.

\vspace{-0.25em}

\section{Conclusion}
\label{sec:conclusion}
\vspace{-0.25em}

In this paper, we introduced SemFlowRAG, a framework that organizes a corpus into a directed semantic hierarchy and guides multi-hop retrieval via a top-down random walk. By enforcing asymmetric directional routing and combining query similarity with a abstractness penalty in the transition weights, SemFlowRAG steers the retrieval process from abstract concepts toward specific evidential passages. Extensive experiments on multiple complex QA benchmarks demonstrate consistent improvements over competitive baselines. Ablation studies confirm that both the directional constraint and the semantic weighting scheme contribute meaningfully to retrieval quality.

\section*{Limitations}

Despite its strong performance, SemFlowRAG has several limitations that warrant discussion. First, the effectiveness of our framework depends on the quality of the underlying semantic graph. The abstractness scores and entity relationships are derived from an offline preprocessing stage; errors or omissions in this stage can propagate downstream and degrade retrieval quality. Second, our top-down directed routing makes an explicit structural assumption—that evidence flows from abstract concepts to specific facts. While this holds broadly across the benchmarks we evaluated, it may not generalize to scenarios requiring lateral reasoning or bottom-up inference, where high-abstractness nodes are the desired targets rather than sources. Third, as shown in our sensitivity analysis, the optimal reset probability varies across datasets, indicating that performance is somewhat sensitive to hyperparameter tuning. Developing an adaptive mechanism to dynamically adjust the exploration depth remains an important direction for future work.

\section*{Ethical Considerations}
\label{sec:ethics}

The development of SemFlowRAG presents ethical considerations common to RAG systems. On the positive side, our framework enhances the transparency and traceability of LLM outputs. By constraining the retrieval trajectory to converge on concrete passage evidence, it helps mitigate LLM hallucinations and allows users to straightforwardly verify the exact evidence used for generation. However, because SemFlowRAG derives its semantic hierarchy and directed edges entirely in a data-driven manner, it is inherently susceptible to the quality of the source corpus. If the corpus contains biases, misinformation, or toxic content, the directed information flow may systematically retrieve and amplify these flawed narratives. Therefore, deploying this framework in real-world, user-facing applications requires careful curation and rigorous auditing of the knowledge base to ensure safety and fairness.

\section*{Acknowledgments}
This work was supported by Shanghai Artificial Intelligence Laboratory.

\bibliography{custom}

\clearpage

\appendix

\section{Appendix}
\label{sec:appendix}

\subsection{Baseline Details}
\label{appendix:baseline_details}
We compare SemFlowRAG with three groups of baselines. First, we include unstructured knowledge representation augmented baselines, covering a no-retrieval setting for QA, the sparse retriever BM25~\cite{robertson1994some} and widely used dense retrievers including Contriever~\cite{izacard2021unsupervised} and GTR~\cite{ni2022large}. Second, we compare against  embedding-based knowledge representation augmented baselines, including GTE-Qwen2-7B-Instruct~\cite{li2023towards}, GritLM-7B~\cite{muennighoff2025generative} and NV-Embed-v2~\cite{lee2025nv}, which represent competitive retrieval-only alternatives based on recent large-scale embedding models. Third, we include structured knowledge representation augmented methods, including RAPTOR~\cite{raptor}, GraphRAG~\cite{han2025rag}, LightRAG~\cite{guo2024lightrag}, HippoRAG~\cite{hipporag} and HippoRAG 2~\cite{hipporag2}.

\subsection{Implementation Details}
\label{appendix:implementation_datails}

\textbf{Backbone models.}
We use Llama-3.3-70B-Instruct as the backbone language model and NV-Embed-v2 as the embedding model. The language model is responsible for OpenIE extraction, fact filtering and final answer generation. NV-Embed-v2 is used to encode passages, entities, facts and queries. We serve Llama-3.3-70B-Instruct with vLLM~\cite{kwon2023efficient} using tensor parallelism on 4 × NVIDIA A100 GPUs.

\textbf{Indexing and OpenIE.}
We build the retrieval graph from OpenIE triples extracted from the corpus. To improve the entity coverage of the constructed graph, we use a high-recall OpenIE prompt that encourages extracting explicit factual statements, including aliases, dates, quantities, locations, affiliations, events, roles and descriptive attributes. We show prompt details in \ref{appendix:kg_build_prompt}. Compound statements are split into separate triples when they express multiple facts. Unless otherwise specified, the maximum generation lengths are 512 tokens for named entity recognition (NER) and 2048 tokens for triple extraction.

\textbf{PPR reset distribution.}
The PPR reset distribution defines where the random walk returns when a reset step is triggered. This distribution is constructed as a mixture of two types of seeds: entity seeds and passage seeds. Entity seeds are obtained from the subject and object entities of the retained query-relevant facts. Their weights are determined by the corresponding query-fact similarity.

Passage seeds are introduced from DPR retrieval scores over passage nodes. All passage nodes are assigned reset weights according to their DPR scores, and these weights are scaled by passage reset weight. In this way, highly relevant passages can receive part of the reset probability directly, while the overall reset distribution is still dominated by entity seeds. During PPR propagation, each reset step returns to this mixed distribution over both entity nodes and passage nodes, rather than returning only to entity seeds.

\textbf{Retrieval configuration.}
We use directed PPR over a graph containing both entity and passage nodes. We set passage reset weight to $0.08$. We use a PPR reset probability of $0.5$. For semantic direction control, the transition probability budget is split into $0.9$ for abstractness decreasing edges and $0.1$ for abstractness increasing edges. We first retrieve the top 12 candidate facts by embedding similarity of query and fact, and then apply an LLM-based fact filter that keeps up to 12 query-relevant facts. We use the entities appearing in the filtered facts to construct the entity part of the PPR reset vector, and retain the top 20 entity seeds.

\textbf{Answer generation.}
For answer generation, the LLM uses the top 5 retrieved passages. We use a QA temperature of $0.4$ and keep the global temperature for other LLM calls at $0$.

\textbf{Efficiency and computational overhead.}
We further analyze the computational and storage overhead introduced by the semantic gradient graph and directed PPR retrieval.
Table~\ref{tab:offline_efficiency} reports the offline construction and storage costs, while Table~\ref{tab:retrieval_latency} reports the average online retrieval latency per query.
The additional storage mainly consists of directed graph metadata, including entity-level abstractness scores, edge directions, and normalized transition weights.
Since this process is performed only once during indexing, the cost can be amortized over subsequent queries.
The additional storage mainly consists of directed graph metadata, including entity-level abstractness scores and transition-related edge information.
During online retrieval, directed PPR introduces additional semantic directional guidance, resulting in moderately higher retrieval latency.

\begin{table}[t]
\centering
\small
\setlength{\tabcolsep}{10pt}
\begin{tabular}{lcc}
\toprule
\textbf{Dataset} &
\textbf{Construction} &
\textbf{Storage (MB)} \\
\midrule
2Wiki       & 34s + 1m41s   & 18.2 $\rightarrow$ 22.5 \\
HotpotQA    & 1m35s + 5m11s & 34.6 $\rightarrow$ 42.9 \\
MuSiQue     & 1m45s + 6m21s & 37.7 $\rightarrow$ 47.5 \\
PopQA       & 1m24s + 4m35s & 31.8 $\rightarrow$ 39.1 \\
NQ          & 1m09s + 6m33s & 31.8 $\rightarrow$ 44.3 \\
LV-Eval     & 6m04s + 1h02m & 96.7 $\rightarrow$ 122.4 \\
NarrativeQA & 3s + 18s      & 5.0 $\rightarrow$ 6.4 \\
\bottomrule
\end{tabular}
\caption{
Offline efficiency comparison between the standard KG retrieval pipeline and SemFlowRAG.
Construction reports the standard KG construction time plus the additional semantic-gradient preprocessing time.
Storage is reported as KG $\rightarrow$ SemFlowRAG.
}
\label{tab:offline_efficiency}
\end{table}

\begin{table}[t]
\centering
\small
\setlength{\tabcolsep}{16pt}
\begin{tabular}{lcc}
\toprule
\textbf{Dataset} &
\textbf{KG} &
\textbf{SemFlowRAG} \\
\midrule
2Wiki       & 1.600 & 2.008 \\
HotpotQA    & 1.775 & 2.579 \\
MuSiQue     & 1.695 & 2.509 \\
PopQA       & 1.508 & 2.247 \\
NQ          & 1.941 & 2.909 \\
LV-Eval     & 2.071 & 3.879 \\
NarrativeQA & 1.314 & 1.342 \\
\bottomrule
\end{tabular}
\caption{
Average online retrieval latency per query in seconds. SemFlowRAG introduces additional directional guidance along the semantic gradient during directed PPR retrieval.
}
\label{tab:retrieval_latency}
\end{table}

\subsection{Additional Experimental Results}

\subsubsection{Sensitivity Analysis}
\label{subsec:sensitivity}

We evaluate the sensitivity of SemFlowRAG to three key hyperparameters: the PPR reset probability, the number of entity seeds and the QA decoding temperature. All experiments are conducted under the default configuration described above, varying only the parameter under test.

\begin{figure}[h]
    \centering
    \includegraphics[width=1\linewidth]{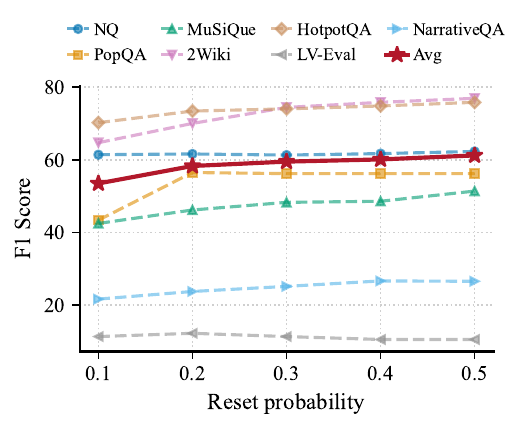}
    \caption{\textbf{Reset probability factor.} We report QA performance (F1 scores) with different PPR reset probabilities in our algorithm.}
    \label{tab:reset_probability}
    \label{fig:Sensitivity_analysis}
\end{figure}

\textbf{PPR reset probability.} This parameter controls how often the random walk returns to the initial seed nodes. As shown in Figure~\ref{fig:Sensitivity_analysis}, increasing it from 0.1 to 0.5 consistently improves the average F1 score from 53.5 to 61.2. A lower reset probability allows deeper graph traversal but risks semantic drift by diluting the probability mass into noisy, irrelevant nodes. The anchoring effect is particularly beneficial for multi-hop datasets: the F1 score on 2Wiki jumps from 64.7 to 76.9, indicating that while complex reasoning requires multi-hop exploration, the necessary evidence typically resides within a few hops of the initial seeds. Although a few datasets (e.g., LV-Eval and PopQA) peak or plateau earlier, a reset probability of 0.5 provides the most robust regularization overall.

\textbf{Entity seeds.} We examine the number of initial entity seeds used in the PPR reset vector (Table~\ref{tab:entity-seed-sensitivity}). For this experiment, the QA temperature is fixed at $0.8$. Varying the seed count from 8 to 20 yields minimal fluctuation in both per-dataset and average F1 scores (64.4-64.5), with a marginal gain of 0.9 points on 2Wiki at 20 seeds. This stability indicates that our fact-filtering pipeline already supplies a sufficient set of query-relevant entities, and that SemFlowRAG is largely insensitive to this parameter within a reasonable range.

\textbf{QA decoding temperature.} We further assess the impact of the QA generation temperature (Table~\ref{tab:qa-temperature-sensitivity}), with the number of entity seeds fixed at $5$. Across the range of 0.0 to 0.8, the average F1 varies within a narrow band of 63.8-64.0, with no dataset exhibiting a monotonic upward or downward trend. This confirms that our approach is robust to decoding stochasticity; a moderate temperature (e.g., 0.2-0.6) offers a good balance between determinism and output diversity, while extreme values incur no substantial degradation.




\begin{table}[t]
\centering
\small
\setlength{\tabcolsep}{3pt}
\begin{tabular}{lcccccc}
\toprule
\textbf{Seeds} & \textbf{NQ} & \textbf{PopQA} & \textbf{MuSiQue} & \textbf{2Wiki} & \textbf{HotpotQA} & \textbf{Avg.} \\
\midrule
8  & 62.6 & 56.0 & 51.6 & 76.0 & 75.7 & 64.4 \\
16 & 62.5 & 56.3 & 51.2 & 76.3 & 75.9 & 64.4 \\
20 & 62.3 & 56.2 & 51.4 & 76.9 & 75.8 & 64.5 \\
\bottomrule
\end{tabular}
\caption{Sensitivity analysis of the number of entity seeds used in the PPR reset vector. All values are QA F1 scores.}
\label{tab:entity-seed-sensitivity}
\end{table}


\subsection{Case Study}
\label{subsec:qualitative}
To illustrate the retrieval and reasoning process of SemFlowRAG, we provide a case study in Figure~\ref{fig:case-study}. The given query requires two-hop reasoning: first identifying the director of the film ``\textit{An Event}'', and subsequently finding the director's child. 

As shown in the retrieved top-5 passages, SemFlowRAG successfully captures the complete evidence chain. Passage 1 resolves the first hop by identifying Vatroslav Mimica as the director of the film. Building upon this, Passage 2 resolves the second hop by explicitly stating that his son is Sergio Mimica-Gezzan. By securing these structurally connected facts at the very top of the retrieved context, SemFlowRAG provides a clear reasoning evidence. Consequently, the LLM is able to accurately generate the right answer.

\begin{figure}[t]
\begin{promptbox}{Case Study: End to End QA Performance of SemFlowRAG}
\textbf{Question:} Who is the child of the director of film \textit{An Event}? \\

\textbf{Ground Truth:} Sergio Mimica-Gezzan \\

\textbf{Retrieved Top-5 Passages:}
\begin{enumerate}
    \item \textbf{An Event:} An Event is a 1969 Yugoslav feature film directed by Vatroslav Mimica, based on a short story by Anton Chekhov...
    \item \textbf{Vatroslav Mimica:} Vatroslav Mimica is a Croatian film director and screenwriter. His son Sergio Mimica-Gezzan is an American film and television director... 
    \item \textbf{M. Night Shyamalan:} Manoj Nelliyattu ``M. Night'' Shyamalan is an American filmmaker and actor...
    \item \textbf{Mario Bava:} Mario Bava was an Italian cinematographer, director, special effects artist and screenwriter...
    \item \textbf{Anup Sengupta:} Anup Sengupta is a Bengali film Director and Producer...
\end{enumerate}

\textbf{Answer:} \textbf{Sergio Mimica-Gezzan.}
\end{promptbox}

\caption{A case study of SemFlowRAG.}
\label{fig:case-study}
\end{figure}












\subsection{Prompt Details}
\subsubsection{KG build prompt}
\label{appendix:kg_build_prompt}

We show LLM prompts for NER and triple extraction in Figure~\ref{fig:ner_prompt} and~\ref{fig:triple_extraction_prompt}, including the system instructions and demonstrations for both tasks. The NER prompt is used to extract named entities from the passage, while the triple extraction prompt further converts the passage and entity list into RDF-style triples.

\begin{figure}[!h]

\begin{promptbox}{NER Prompt}
\textbf{System prompt:}

Your task is to extract named entities from the given paragraph. 
Respond with a JSON list of entities.

\vspace{0.5em}
\textbf{Demonstration input:}

Radio City

Radio City is India's first private FM radio station and was started on 3 July 2001.
It plays Hindi, English and regional songs.

Radio City recently forayed into New Media in May 2008 with the launch of a music portal - PlanetRadiocity.com that offers music related news, videos, songs, and other music-related features.

\vspace{0.5em}
\textbf{Demonstration output:}

\{``named\_entities'':

\hspace*{1em}[``Radio City'', ``India'', ``3 July 2001'', ``Hindi'', ``English'', ``May 2008'', ``PlanetRadiocity.com'']

\}
\end{promptbox}
\caption{Prompts and demonstrations for NER}
\label{fig:ner_prompt}
\end{figure}

\begin{figure}[hbtp]
\begin{promptbox}{Triple Extraction Prompt}
\textbf{System prompt:}

Your task is to construct an RDF (Resource Description Framework) graph from the given passages and named entity lists.
Respond with a JSON list of triples, with each triple representing a relationship in the RDF graph.

Pay attention to the following requirements:
\begin{itemize}
    \setlength\itemsep{0em} 
    \item Extract as many explicit factual statements as possible from the paragraph; aim for high recall rather than a small summary.
    \item Each triple should contain at least one named entity when possible, but do not omit explicit facts whose useful endpoints are descriptive phrases, dates, quantities, titles, roles, events, locations, affiliations, or other salient noun phrases.
    \item Split enumerations and compound statements into separate triples when they express multiple facts.
    \item Preserve factual specificity from the paragraph, including aliases, dates, numbers, locations, occupations, memberships, affiliations, works, events, and descriptive attributes.
    \item Do not infer facts that are not explicitly supported by the paragraph.
    \item Clearly resolve pronouns to their specific names to maintain clarity.
\end{itemize}

\vspace{0.5em}
\textbf{Demonstration input:}

Convert the paragraph into a JSON dict, it has a named entity list and a triple list.

Paragraph: 

Radio City 

Radio City is India's first private FM radio station and was started on 3 July 2001. It plays Hindi, English and regional songs. 

Radio City recently forayed into New Media in May 2008 with the launch of a music portal - PlanetRadiocity.com that offers music related news, videos, songs, and other music-related features.

\{``named\_entities'':

\hspace*{1em}[``Radio City'', ``India'', ``3 July 2001'', ``Hindi'', ``English'', ``May 2008'', ``PlanetRadiocity.com'']

\}

\vspace{0.5em}
\textbf{Demonstration output:}

\{``triples'': [

\hspace*{1em}[``Radio City'', ``located in'', ``India''],\\
\hspace*{1em}[``Radio City'', ``is'', ``private FM radio station''],\\
\hspace*{1em}[``Radio City'', ``started on'', ``3 July 2001''],\\
\hspace*{1em}[``Radio City'', ``plays songs in'', ``Hindi''],\\
\hspace*{1em}[``Radio City'', ``plays songs in'', ``English''],\\
\hspace*{1em}[``Radio City'', ``forayed into'', ``New Media''],\\
\hspace*{1em}[``Radio City'', ``launched'', ``PlanetRadiocity.com''],\\
\hspace*{1em}[``PlanetRadiocity.com'', ``launched in'', ``May 2008''],\\
\hspace*{1em}[``PlanetRadiocity.com'', ``is'', ``music portal''],\\
\hspace*{1em}[``PlanetRadiocity.com'', ``offers'', ``news''],\\
\hspace*{1em}[``PlanetRadiocity.com'', ``offers'', ``videos''],\\
\hspace*{1em}[``PlanetRadiocity.com'', ``offers'', ``songs'']

\hspace*{1em}]
\}
\end{promptbox}
\caption{Prompts and demonstrations for triple extraction}
\label{fig:triple_extraction_prompt}
\end{figure}

\subsubsection{LLM fact filter prompt}

We show LLM prompts for fact filtering in Figure~\ref{fig:fact_filter_prompt}. The prompt requires the model to select relevant facts from candidate triples while preserving answer-bearing facts and bridge facts for multi-hop reasoning.

\begin{table}[!h]
\centering
\small
\resizebox{\columnwidth}{!}{
\begin{tabular}{lccccc}
\toprule
\textbf{Dataset} & \textbf{0.0} & \textbf{0.2} & \textbf{0.4} & \textbf{0.6} & \textbf{0.8} \\
\midrule
NQ       & 0.6235 & 0.6264 & 0.6267 & 0.6290 & 0.6193 \\
PopQA    & 0.5573 & 0.5588 & 0.5587 & 0.5598 & 0.5639 \\
MuSiQue  & 0.5138 & 0.5072 & 0.5023 & 0.5037 & 0.5015 \\
2Wiki    & 0.7445 & 0.7536 & 0.7602 & 0.7526 & 0.7542 \\
HotpotQA & 0.7525 & 0.7531 & 0.7541 & 0.7540 & 0.7566 \\
Avg.     & 0.6383 & 0.6398 & 0.6404 & 0.6398 & 0.6391 \\
\bottomrule
\end{tabular}
}
\caption{Sensitivity analysis of QA decoding temperature. All values are QA F1 scores.}
\label{tab:qa-temperature-sensitivity}
\end{table}

\begin{figure*}

\begin{promptbox}{Fact Filtering}
\textbf{System prompt:} \\
Your input fields are: question, which denotes the query for retrieval, and fact\_before\_filter, which denotes the candidate facts to be filtered. The output field is fact\_after\_filter, which stores the filtered facts in JSON format. All interactions follow the structure: question, fact\_before\_filter, and fact\_after\_filter. The output must be parseable according to the JSON schema:
\{``fact'': [[subject, predicate, object], ...]\}

The model should select up to 12 facts that are most relevant to the question. Prefer keeping complementary bridge facts, entity-linking facts, and answer-bearing facts when they may support multi-hop reasoning. If no candidate fact is relevant, it should return \{``fact'': []\}.
The model must only use facts from the candidate list and must not generate new facts.
\vspace{1em}

\textbf{Demonstration 1:} 

\textbf{Question:} Are Imperial River (Florida) and Amaradia (Dolj) both located in the same country?

\textbf{Fact Before Filter:}
\{``fact'': [\\
\hspace*{1em}[``imperial river'', ``is located in'', ``florida''],\\
\hspace*{1em}[``imperial river'', ``is a river in'', ``united states''],\\
\hspace*{1em}[``imperial river'', ``may refer to'', ``south america''],\\
\hspace*{1em}[``amaradia'', ``flows through'', ``roia de amaradia''],\\
\hspace*{1em}[``imperial river'', ``may refer to'', ``united states'']\\
]\}

\textbf{Fact After Filter:}
\{``fact'': [\\
\hspace*{1em}[``imperial river'', ``is located in'', ``florida''],\\
\hspace*{1em}[``imperial river'', ``is a river in'', ``united states''],\\
\hspace*{1em}[``amaradia'', ``flows through'', ``roia de amaradia'']\\
]\}

\vspace{0.5em}

\textbf{Demonstration 2:} \\
\textbf{Question:} When is the director of film The Ancestor's birthday? \\
\textbf{Fact Before Filter:}
\{``fact'': [\\
\hspace*{1em}[``jean jacques annaud'', ``born on'', ``1 october 1943''],\\
\hspace*{1em}[``sui hark'', ``born on'', ``15 february 1950''],\\
\hspace*{1em}[``pablo trapero'', ``born on'', ``4 october 1971''],\\
\hspace*{1em}[``the ancestor'', ``directed by'', ``guido brignone''],\\
\hspace*{1em}[``benh zeitlin'', ``born on'', ``october 14 1982'']\\
]\}

\textbf{Fact After Filter:} \{``fact'': [[``the ancestor'', ``directed by'', ``guido brignone'']]\}
\end{promptbox}
\caption{Fact filtering prompt and demonstrations}

\label{fig:fact_filter_prompt}
    
\end{figure*}

\end{document}